# Investigation of Deep Neural Network Acoustic Modelling Approaches for Low Resource Accented Mandarin Speech Recognition

# 关于深度神经网络建模方法

# 用于数据缺乏的带口音普通话语音识别的研究

*Xurong Xie* (谢旭荣) [1,3], *Xiang Sui* (隋相) [1,3], *Xunying Liu* (刘循英) [1,2] *& Lan Wang* (王岚) [1,3]

[1] Key Laboratory of Human-Machine Intelligence-Synergy Systems,
Shenzhen Institutes of Advanced Technology, Chinese Academy of Sciences, China
[2] Cambridge University Engineering Dept, Trumpington St., Cambridge, CB2 1PZ U.K.
[3] The Chinese University of Hong Kong, Hong Kong, China
[1] 人机智能协同系统重点实验室，中国科学院深圳先进技术研究院，中国
[2] 剑桥大学工程系，剑桥，英国
[3] 香港中文大学，香港，中国

xr.xie@siat.ac.cn, xiang.sui@siat.ac.cn, xl207@cam.ac.uk, lan.wang@siat.ac.cn

## Abstract

The Mandarin Chinese language is known to be strongly influenced by a rich set of regional accents, while Mandarin speech with each accent is quite low resource. Hence, an important task in Mandarin speech recognition is to appropriately model the acoustic variabilities imposed by accents. In this paper, an investigation of implicit and explicit use of accent information on a range of deep neural network (DNN) based acoustic modelling techniques is conducted. Meanwhile, approaches of multi-accent modelling including multi-style training, multi-accent decision tree state tying, DNN tandem and multi-level adaptive network (MLAN) tandem hidden Markov model (HMM) modelling are combined and compared in this paper. On a low resource accented Mandarin speech recognition task consisting of four regional accents, an improved MLAN tandem HMM systems explicitly leveraging the accent information was proposed and significantly outperformed the baseline accent independent DNN tandem systems by 0.8%-1.5% absolute (6%-9% relative) in character error rate after sequence level discriminative training and adaptation.

## 摘要

众所周知中文普通话被众多的地区口音强烈地影响着，然而带每个口音的普通话语音数据却是十分缺乏。因此，普通话语音识别的一个重要目标是恰当地模拟口音带来的声学变化。本文给出了隐式和显式地使用口音信息的一系列基于深度神经网络（DNN）的声学模型技术的研究。与此同时，包括混合条件训练，多口音决策树状态绑定，DNN 级联和多级自适应网络（MLAN）级联隐马尔可夫模型（HMM）建模等的多口音建模方法在本文中被组合和比较。一个能显式地利用口音信息的改进的 MLAN 级联 HMM 系统被提出并应用于一个由四个地区的口音组成的数据缺乏的带口音普通话语音识别任务中。在经过序列区分性训练和自适应后，通过绝对上 0.8%到 1.5%（相对上 6%到 9%）的字错误率下降，该系统显著地优于基线的口音独立 DNN 级联系统。

## 1. Introduction

An important part of the Mandarin speech recognition task is to appropriately handle the influence from a rich set of diverse accents. There are at least seven major regional accents in China [1, 2]. The related variabilities imposed on accented Mandarin speech are complex and widespread. The resulting high mismatch can lead to severe performance degradation for automatic speech recognition (ASR) tasks. To handle this problem, ASR systems can be trained on large amounts of accent specific speech data [3]. However, collecting and annotating accented data is very expensive and time-consuming. Hence, the amount of available accent specific speech data is often quite limited.

An alternative approach is to exploit the accent independent features among standard Mandarin speech data, which are often available in large amounts, to improve robustness and generalization. Along this line, two categories of techniques can be used. The first category of techniques aim to directly adapt systems trained on standard Mandarin speech data [4, 5, 6, 7, 8]. The second category uses standard Mandarin speech to augment the limited in-domain accent specific data in a multi-style training framework [9]. For example, an accent dependent phonetic decision tree tying technique was proposed in [10, 11]. It allows the resulting acoustic models to explicitly learn both the accent independent and the accent specific characteristics in speech.

Recently deep neural networks (DNNs) have become increasing popular for acoustic modelling, due to their inherent robustness to the highly complex factors of variabilities found in natural speech [12, 14, 15]. These include external factors such as environment noise [16. 17. 18], and language dependent linguistic features [19, 20, 21]. In order to incorporate DNNs, or multi-layer perceptrons (MLPs) in general, into hidden Markov model (HMM)-based acoustic models, two approaches can be used. The first uses a hybrid architecture that estimates the HMM state emission probabilities using DNNs [22]. The second approach uses an MLP or DNN [17], which is trained to produce phoneme posterior probabilities, as a feature extractor. The resulting probabilistic features [23] or bottleneck features [24], are concatenated with standard front-ends and used to train Gaussian mixture model (GMM)-HMMs in a tandem fashion. As GMM-HMMs remain as the back-end classifier, the tandem approach requires minimum change to the downstream techniques, such as adaptation and discriminative training, while retaining the useful information by the bottleneck features.

Using limited amounts of accented data alone is insufficient to obtain sufficient generalization for the resulting acoustic models, including DNNs. Therefore, a key problem in accented Mandarin speech recognition with low resources, as considered in this paper, is how to improve coverage and generalisation by exploiting the commonalities and specialties among standard and accented speech data during training. Using conventional multi-style DNN training based on a mix of standard and accented Mandarin speech data, accent independent features found in both can be implicitly learned [25, 26].

Inspired by recent works on multi-lingual low resource speech recognition [19, 27, 20, 21], this paper aims to investigate and compare the explicit as well as the implicit uses of accent information in state-of-the-art deep neural network (DNN) based acoustic modelling techniques, including conventional tied state GMM-HMMs, DNN tandem systems, and multi-level adaptive network (MLAN) [27, 28] tandem HMMs. These approaches are evaluated on a low resource accented Mandarin speech recognition task consisting of accented speech collected from four regions: Guangzhou, Chongqing, Shanghai and Xiamen. The improved multi-accent GMM-HMM and MLAN tandem systems explicitly leveraging the accent information during model training significantly outperformed the baseline GMM-HMM and DNN tandem HMM systems by 0.8%-1.5% absolute (6%-9% relative) in character error rate after minimum phone error (MPE) based discriminative training and adaptation.

The rest of this paper is organized as follows. Standard acoustic accent modelling approaches are reviewed in section 2. These include multi-accent decision tree state tying for GMM-HMM systems, and multi-accent DNN tandem systems. MLAN tandem systems with improved pre-training for accent modelling are presented in section 3.2. Experimental results are presented in section 4. Section 5 draws the conclusions and discusses future work.

# 2. Acoustic modelling for accented speech

## 2.1. Multi-style accent modelling

Multi-style training [9] is used in this paper for accent modelling. This approach uses speech data collected in a wide range of styles and domains. Then, it exploits the implicit modelling ability of mixture models used in GMM-HMMs and, more recently, deep neural networks [16, 20, 21] to obtain a good generalization to unseen situations. In the accented speech modelling experiments of this paper, large amount of standard Mandarin speech data is used to augment the limited accented data during training to provide useful accent independent features.

## 2.2. Multi-accent decision tree state tying

As the phonetic and phonological realization of Mandarin speech is significantly different between regional accents, inappropriate tying of context dependent HMM states associated with different accents can lead to poor coverage and discrimination for GMM-HMM based acoustic models. In order to handle this problem, decision tree clustering [10, 11] with multi-accent branches is tried in this paper. In order to effectively exploit the commonalities and specificities found in standard and accented Mandarin data, accent dependent (AD) questions are used together with conventional linguistic questions during the clustering process. A sample of the accented branches is shown in red part of figure 1.

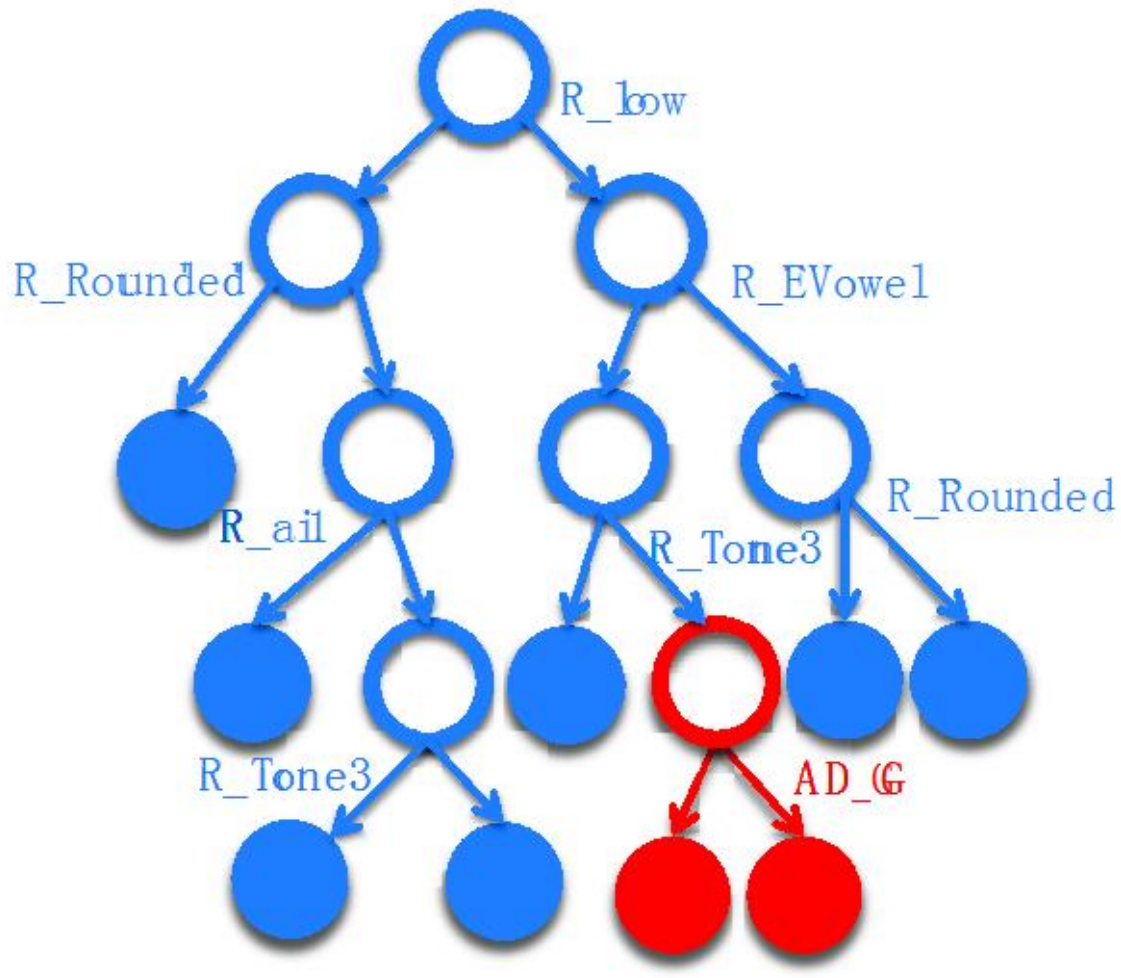

Figure 1: A part of multi-accent decision tree. Blue: conventional branches; Red: accented branches

In common with standard maximum likelihood (ML) based phonetic decision tree tying [13], the questions giving highest log-likelihood improvement are chosen when splitting tree nodes. The algorithm iterates until no more splitting operations can yield a log-likelihood increase above a certain threshold. Therefore, the multi-accent information is explicitly used during states tying. As expected, the use of accent dependent questions dramatically increases the number of context-dependent phone units to consider during training and decoding. As not all of them are allowed by the lexicon, following the approach proposed in [29], only the valid subset under the lexical constraint is considered in this paper.

### 2.3. Multi-accent DNN tandem systems

In this paper, DNNs are trained to extract bottleneck features to be used in both DNN tandem and MLAN tandem systems. They are trained to model frame posterior probabilities of context-dependent phone HMM state targets. The inputs to DNNs consist of a context window of 11 successive frames of features for each time instance. The input to each neuron of each hidden layer is a linearly weighted sum of the outputs from the previous layer, before fed into a sigmoid activation function. At the output layer a softmax activation is used to compute posterior probability of each output target. The networks were first pre-trained by layer-by-layer restricted Boltzmann machine (RBM) pre-training [14, 15], then globally fine-tuned to minimize the frame-level cross-entropy by back-propagation. Moreover, the last hidden layer is set to have a significantly smaller number of neurons [24]. This narrow layer introduces a constriction in dimensionality while retaining the information useful for classification. Subsequently, low dimensional bottleneck features can be extracted by taking neuron values of this layer before activation. The bottleneck features are then appended to the standard acoustic features and used to train the back-end GMM-HMMs in tandem systems.

## 3. Multi-accent MLAN tandem systems

### 3.1. Multi-level adaptive network tandem systems

A multi-level adaptive network (MLAN) was first proposed for cross domain adaptation [27, 28], where large amounts of out-of-domain telephone and meeting room speech were used to improve the performance of an ASR system trained on a limited amount of in-domain multi-genre archive broadcast data. The MLAN approach explored the useful domain independent characteristics in the out-of-domain data to improve in-domain modelling performance, while reducing the mismatch across different domains. In this paper, the MLAN approach is further exploited to improve the performance of accented Mandarin speech recognition systems.

An MLAN system consists of two component subnetworks. The first-level network is trained with acoustic features of large amounts of accent independent standard Mandarin speech data. The acoustic features of target accented speech data are then fed forward through the first-level network. The resulting bottleneck features are then concatenated with the associated standard acoustic features and used as input to train the second-level network. After both of the two component networks are trained, the entire training set, including both standard and accented Mandarin speech data, is fed forward through the two subnetworks in turn. The resulting set of bottleneck features are then concatenated with the standard front-ends and used to train the back-end GMM-HMMs.

### 3.2. Improved MLAN tandem systems for accent modelling

The MLAN framework can be considered as stacked DNNs that consists of multi level of networks [21, 20]. The second level network of stacked DNNs uses the information of first level network only in the input features, while weights and biases in the second level network are randomly initialized before pre-training and training. One important issue associated with conventional MLAN systems is the robust estimation of the second level DNN parameters. When using limited amounts of in-domain, accent specific speech data to adapt the second level DNN, as is considered in this work, a direct update of its associated weight parameters presents a significant data sparsity problem and can lead to poor generalization performance [21, 25, 26]. In order to address this issue, an improvement form of pre-training initialization is used in this paper for the second level DNN.

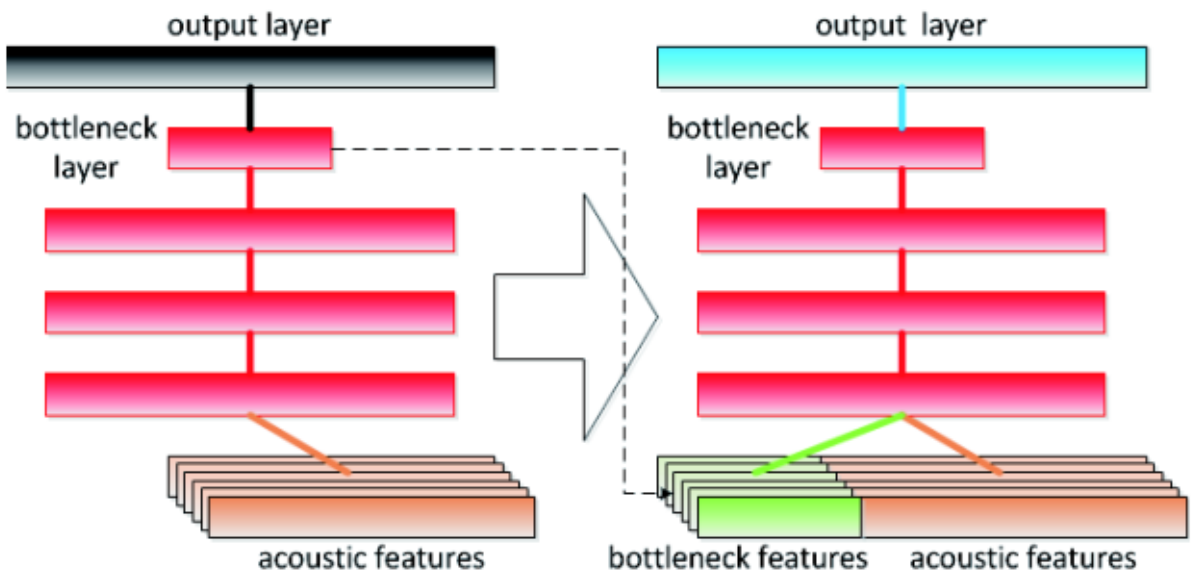

Figure 2: Improved MLAN training for tandem systems. Left: first level DNN network; Right: second level DNN network.

First, all the hidden layers parameters of the second level accent adaptive DNN, and its input layer parameters associated with the standard acoustic features (show as red and orange parts in figure 2) are initialized using those of the first level DNN trained on sufficient amounts of accent independent speech data. Second, the remaining input layer weights and biases connecting the input bottleneck features generated from first level DNN are initialized using RBM pre-training (shown as green in figure 2).

When training the second level DNN, the parameters between the bottleneck layer and the output layer are updated first (shown as blue in figure 2), while fixing the rest of the second level network. The entire second level network is then globally fine-tuned using back-propagation. Similar to the multilingual DNN adaptation approach investigated in [21], the proposed method aims to adapt the second level network parameters based on those of a well trained first level network.

## 4. Experiments and results

### 4.1. Data description

In this section the performance of various accented Mandarin speech recognition approaches are evaluated. 43 hours of standard Mandarin speech [30] and 22.6 hours of accented Mandarin speech containing Guangzhou, Chongqing, Shanghai and Xiamen regional accents [31] released by CASIA and RASC863 databases respectively were used in training. Four testing sets associated with each of these four accents respectively were also used. More detailed information of these data sets are presented in table 1.

| Data Source | Database | Train (hrs) | Test (hrs) |
|---|---|---|---|
| Standard (st) Mandarin | CASIA | 42.9 | - |
| Guangzhou (gz) accent | RASC863 | 6.0 | 1.7 |
| Chongqing (cq) accent | RASC863 | 6.0 | 1.6 |
| Shanghai (sh) accent | RASC863 | 5.5 | 1.4 |
| Xiamen (xm) accent | RASC863 | 5.4 | 1.5 |

Table 1: Standard and accented Mandarin speech data sets

### 4.2. Experiment setup

Baseline context-dependent phonetic decision tree clustered [13, 32] triphone GMM-HMM systems with 16 Gaussians per state were trained using 42 dimensional acoustic features consisting of heteroskedastic linear discriminant analysis (HLDA) perceptual linear predictive (PLP) features and pitch parameters. These were used as the input features, and to produce accent independent state level alignment to train DNNs with 2048 neurons in each non-bottleneck hidden layer using the Kaldi toolkit [33]. Meanwhile the bottleneck layer had 26 neurons. All DNNs were trained with initial learning rate of 0.008 and the commonly used newbob annealing schedule. Mean normalization and principle component analysis (PCA) de-correlation were applied to the resulting bottleneck features before being appended to the above acoustic features.

### 4.3. Performance of multi-accent GMM-HMM systems

The performance of multi-accent GMM-HMM systems were first evaluated on Guangzhou accented speech data. These are shown in table 2. In this table, the “Model AD” column denotes accent dependent questions were used in decision tree state tying. This table shows that the multi-accent HMM model (Sys (2)) trained by adding all four types of accented speech to the standard Mandarin data outperformed folding in Guangzhou accent data only (Sys (1)). In addition, the explicit use of accent information during decision tree clustering (Sys (3)) obtained a further character error rate (CER) reduction of 2.7% absolute from 17.77% down to 15.07%.

| Sys | Model | Model AD | Baseline trn | CER (%) |
|---|---|---|---|---|
| (1) | Baseline HMM | × | st + gz | 20.06 |
| (2) | | | st + all | 17.77 |
| (3) | | √ | | **15.07** |

Table 2: Performance of baseline GMM-HMM systems trained on standard Mandarin speech plus Guangzhou accent data only, or all four accents of table 1, and evaluated on Guangzhou accent test set.

### 4.4. Performance of multi-accent DNN tandem systems

A second set of experiments comparable to those shown in table 2 were then conducted to evaluate the performance of four tandem systems on the Guangzhou accent test set. In addition to the standard Mandarin speech data, the Guangzhou accent data, or all 4 accent types, were also used in DNN training. All DNNs here had 4 hidden layers including the bottleneck layer. These are shown in table 3. The multi-accent trained DNN tandem system (Sys (4) in table 3, which used both accent dependent questions in decision tree based HMM state clustering, and included all four accent types in DNN training, gave the lowest character error rate of 13.16%.

| Sys | Model | Model AD | Baseline trn | DNN trn | CER (%) |
|---|---|---|---|---|---|
| (1) | DNN tandem | × | st + gz | st | 17.14 |
| (2) | | | | st + gz | 15.85 |
| (3) | | | st + all | st + all | 14.12 |
| (4) | | √ | | | **13.16** |

Table 3: Performance of DNN tandem systems on Guangzhou accent test set

### 4.5. Performance of multi-accent MLAN tandem systems

The performance of various MLAN tandem systems on Guangzhou accent speech data are shown in table 4. In addition to the standard Mandarin speech data, all 4 accent types were used in both baseline HMM and the first level DNN training. The first level DNN had 4 hidden layers. The first four MLAN tandem systems used a conventional random initialization of the second level DNN with 2 or 4 hidden layers prior to pre-training and full network update on the target accent data. As discussed in sections 1 and 3.2, when using limited amounts of accent specific speech data to estimate the second level DNN, a direct update its associated weight parameters can lead to unrobust estimation and poor generalization. This is shown in the first four lines of table 4. Increasing the number of hidden layers from 2 to 4 for the second level DNN led to further performance degradation. Compared with the best DNN tandem system shown in the bottom line of table 3, a performance degradation of 0.92% absolute was observed.

| Sys | Model | 2nd DNN | | Model AD | CER (%) |
|---|---|---|---|---|---|
| | | Initial | Hidden | | |
| (1) | MLAN tandem | random | 2 | × | 14.35 |
| (2) | | | 4 | | 15.42 |
| (3) | | | 2 | √ | 13.24 |
| (4) | | | 4 | | 14.08 |
| (5) | | 1st DNN | 4 | × | 14.00 |
| (6) | | | | √ | **12.96** |

Table 4: Performance of MLAN tandem systems on Guangzhou accent test set

In contrast, when the improved pre-training based MLAN tandem system discussed in section 3.2 was used, as is shown in the last two lines in table 4, consistent improvements were obtained using both the accent independent and dependent MLAN tandem configurations over the comparable DNN tandem systems shown in the last two line of table 3.

### 4.6. Performance evaluation on multiple accent test sets

A full set of experiments were finally conducted to evaluate the performance of various multi-accent systems on four accent test sets: Guangzhou, Chongqing, Shanghai and Xiamen. The performance of these systems are presented in table 5 and table 6 for the multi-accent GMM-HMM, DNN tandem and improved MLAN tandem systems. “+MPE” denotes MPE discriminative training [34] performed on the maximum likelihood trained “ML” model, and “+MLLR” denotes a subsequent maximum likelihood linear regression (MLLR) adaptation [35] on the “+MPE” model. Moreover, System (0) used only out of domain data, namely standard Mandarin data, to train the GMM-HMMs, which denoted by “$\text{HMM}^{(out)}$”. Meanwhile, “$\text{HMM}^{(ma)}$” denotes multi-accent GMM-HMM systems trained with all accented data as well as standard Mandarin data. Both the DNN tandem and improved MLAN tandem systems utilized the “$\text{HMM}^{(ma)}$ ML” models as their baselines. All DNNs here had 6 hidden layers including the bottleneck layer. “DNN AD” denotes DNN trained with accent dependent state alignment, while all DNNs used in MLAN tandem systems were trained with accent independent state alignment.

| Sys | Model | Model AD | DNN AD | Back -end | CER (%) | | | | |
|---|---|---|---|---|---|---|---|---|---|
| | | | | | gz | cq | sh | xm | Avg. |
| (0) | HMM$^{(out)}$ | × | × | ML | 27.65 | 29.78 | 33.64 | 37.98 | 32.26 |
| (1) | HMM$^{(ma)}$ | × | × | ML | 17.77 | 20.02 | 21.57 | 22.59 | 20.49 |
| | | | | +MPE | 15.30 | 17.36 | 20.12 | 20.87 | 18.41 |
| | | | | +MLLR | 13.75 | 15.48 | 17.81 | 18.94 | 16.50 |
| (2) | | √ | × | ML | 15.07 | 18.11 | 20.71 | 21.88 | 18.94 |
| | | | | +MPE | 12.87 | 15.48 | 18.70 | 19.41 | 16.62 |
| | | | | +MLLR | 11.97 | 13.92 | 16.56 | 17.60 | **15.01** |
| (3) | DNN tandem | × | × | ML | 13.44 | 15.87 | 19.13 | 18.05 | 16.62 |
| | | | | +MPE | 12.71 | 14.79 | 18.37 | 16.97 | 15.71 |
| | | | | +MLLR | 11.91 | 13.77 | 16.38 | 16.09 | 14.54 |
| (4) | | √ | × | ML | 12.83 | 15.00 | 18.79 | 17.67 | 16.07 |
| | | | | +MPE | 12.16 | 13.91 | 17.92 | 16.49 | 15.12 |
| | | | | +MLLR | 11.45 | 12.83 | 16.28 | 15.54 | **14.03** |
| (5) | | √ | √ | ML | 12.94 | 15.17 | 18.78 | 17.57 | 16.12 |
| | | | | +MPE | 12.25 | 13.94 | 17.94 | 16.75 | 15.22 |
| | | | | +MLLR | 11.36 | 12.86 | 16.20 | 15.72 | 14.04 |

Table 5: Performance of baseline multi-accent GMM-HMM and DNN tandem systems evaluated on all four accent test sets

A general trend can be found in tables 5 and 6 that the explicit use of accent information in training lead to consistent improvements for GMM-HMM, DNN tandem and MLAN tandem systems. For example, by explicitly using accent information during model training, an absolute CER reduction of 1.5% (relative 9%) was obtained on the GMM-HMM systems (Sys (2) compared to Sys (1) in table 5). Although the improved MLAN tandem systems got less improvement from MPE training than the DNN tandem systems, they got more significant amelioration when MLLR adaptation was utilized. This indicates that the improved MLAN framework is not exclusive to the MLLR adaptation. The best performance was obtained using the improved MLAN tandem system with accent dependent modelling (Sys (4) in table 6). Using this improved MLAN tandem system, an average CER reduction of 0.8% absolute (6% relative ) was obtained over the baseline DNN tandem system trained without explicitly using any accent information (Sys (3) in table 5).

| Sys | Model | Model AD | 1st DNN trn | Back -end | CER (%) | | | | |
|---|---|---|---|---|---|---|---|---|---|
| | | | | | gz | cq | sh | xm | Avg. |
| (1) | MLAN tandem | × | st | ML | 13.79 | 15.96 | 19.14 | 18.55 | 16.86 |
| | | | | +MPE | 13.36 | 15.37 | 18.87 | 18.14 | 16.44 |
| | | | | +MLLR | 12.20 | 13.98 | 16.71 | 16.57 | 14.87 |
| (2) | | | st + all | ML | 13.39 | 15.14 | 18.66 | 17.78 | 16.24 |
| | | | | +MPE | 13.10 | 14.57 | 18.32 | 17.26 | 15.81 |
| | | | | +MLLR | 11.96 | 13.19 | 16.17 | 15.85 | 14.29 |
| (3) | | √ | st | ML | 12.74 | 15.01 | 18.85 | 17.84 | 16.11 |
| | | | | +MPE | 12.50 | 14.62 | 18.63 | 17.39 | 15.79 |
| | | | | +MLLR | 11.58 | 12.98 | 16.40 | 15.80 | 14.19 |
| (4) | | | st + all | ML | 12.56 | 14.53 | 18.12 | 17.29 | 15.63 |
| | | | | +MPE | 12.20 | 14.04 | 17.94 | 16.86 | 15.26 |
| | | | | +MLLR | 11.10 | 12.55 | 15.70 | 15.46 | **13.70** |

Table 6: Performance of improved MLAN tandem systems evaluated on all four accent test sets

Comparing the results to previous works evaluated also on RASC863 database, researches [36] and [37] used the augmented HMM and dynamic Gaussian mixture selection (DGMS), instead of multi-style accent modelling HMM and multi-accent decision tree state tying used in this paper. Their error rate for Guangzhou (Yue), Chongqing (Chuan) and Shanghai (Wu) accented Mandarin ASR stayed above 40% in syllable level (SER), and the best relative SER reduction against HMM trained with standard Mandarin (Putonghua) was about 20%. Although SER is not directly comparable to CER, but can still be seen as a reference. Meanwhile, for these three accents the comparable HMM system in this paper (Sys (2) in table 5) obtained ML CER of about 18%, which had relative reduction of more than 40% against system (0) in table 5. It might be because that information of standard Mandarin cannot complement the low resource accented Mandarin in the augmented HMM and DGMS approaches.

# 5. Conclusions

In this paper implicit and explicit accent modeling approaches were investigated for low resource accented Mandarin speech recognition. The improved multi-accent GMM-HMM and MLAN tandem systems significantly outperformed the baseline GMM-HMM and DNN tandem HMM systems by 0.8%-1.5% absolute (6%-9% relative) in character error rate after MPE training and adaptation. Experimental results suggest the proposed techniques may be useful for accented speech recognition. Future work will focus on modelling a larger and more diverse set of accents.

# 6. Acknowledgements

This work is supported by National Natural Science Foundation of China (NSFC 61135003), Shenzhen Fundamental Research Program (JCYJ20130401170306806, JC201005280621A).